%% file: main.tex
\documentclass[10pt,twocolumn,letterpaper]{article}

\usepackage[pagenumbers]{cvpr} 

\input{preamble}

\definecolor{cvprblue}{rgb}{0.21,0.49,0.74}
\usepackage[pagebackref,breaklinks,colorlinks,citecolor=cvprblue]{hyperref}

\def\paperID{27} 
\def\confName{CVPR}
\def\confYear{2024}

\title{Comparing fine-grained and coarse-grained object detection for ecology}

\author{Jess Tam\\
University of New South Wales, Sydney\\
{\tt\small j.tam@unsw.edu.au}
\and
Justin Kay\\
Massachusetts Institute of Technology\\
{\tt\small kayj@mit.edu}
}

\begin{document}
\maketitle
\input{sec/0_abstract}
\input{sec/1_intro}
\input{sec/2_related_works}
\input{sec/3_methods}
\input{sec/4_results}
\input{sec/5_discussion}
\input{sec/6_acknowledgement}
{
    \small
    \bibliographystyle{ieeenat_fullname}
    \bibliography{main}
}


\end{document}

%% file: preamble.tex
\usepackage[dvipsnames]{xcolor}


%% file: sec/0_abstract.tex
\begin{abstract}

Computer vision applications are increasingly popular for wildlife monitoring tasks. While some studies focus on the monitoring of a single species, such as a particular endangered species, others monitor larger functional groups, such as predators. In our study, we used camera trap images collected in north-western New South Wales, Australia, to investigate how model results were affected by combining multiple species in single classes, and whether the addition of negative samples can improve model performance. We found that species that benefited the most from merging into a single class were mainly species that look alike morphologically, i.e. macropods. Whereas species that looked distinctively different gave mixed results when merged, e.g. merging pigs and goats together as non-native large mammals. We also found that adding negative samples improved model performance marginally in most instances, and recommend conducting a more comprehensive study to explore whether the marginal gains were random or consistent. We suggest that practitioners could classify morphologically similar species together as a functional group or higher taxonomic group to draw ecological inferences. Nevertheless, whether to merge classes or not will depend on the ecological question to be explored.

\end{abstract}

%% file: sec/1_intro.tex
\section{Introduction}
\label{sec:intro}

There are many avenues to monitor ecosystem health, including estimating the population density of wildlife, behavioural monitoring, and community diversity \cite{glover2019camera}. Traditionally, population density is determined by methods such as capture-mark-recapture \cite{schwarz1996general}. Recently, motion-triggered camera traps have enabled mass data collection while reducing the manual labour involved in capture-mark-recapture, thus encouraging ecologists to collect data remotely \cite{delisle2021next}.

Analysing camera trap data is often labour-intensive because of the vast amounts of imagery that is captured by the cameras. An appealing option is using computer vision to automate some or all of this processing. 
A number of recent developments have made this increasingly practical for ecologists, including highly efficient and accurate object detection networks that can be used to localize and classify wildlife in imagery~\cite{szegedy2015goingdeep,redmon2015you,terven2023comprehensive}. In particular, the MegaDetector~\cite{beery2019efficient}---a pre-trained YOLOv5-based model for wildlife detection---has made ecological computer vision processing widely accessible. The MegaDetector classifies images into coarse categories of 'animal', 'vehicle', 'person', and 'empty' \cite{beery2019efficient}. This is a useful first step in data processing, however, in order to make ecological inferences, ecologists must build additional species-level models.

Fine-grained classification is a known challenge in computer vision~\cite{van2018inaturalist,wei2021fine,zhao2017survey,wei2019deep}. Unfortunately many classification problems in ecology are fine-grained, with many species exhibiting highly similar physical characteristics. On the other hand, this level of granularity is not always necessary for ecological analysis. For instance, ecologists may be interested in the behavior of \textit{functional groups} of species that share similar ecosystem functions, where the differentiation of species within groups may be of less concern.


In this paper we investigate the challenges of training computer vision models for wildlife detection and classification from an ecological perspective. In particular, we ask: \textbf{(1)} How does the level of granularity of species recognition models affect ecological analysis? We construct a dataset to study this question (\cref{sec:methods}) by comparing model performance on both fine-grained (\ie species-level) and coarse-grained (\ie functional group-level) categories. \textbf{(2)} How do negative samples---data that do not contain target objects, which are plentiful in camera traps due to false triggers---affect model accuracy? 



%% file: sec/2_related_works.tex
\section{Related works}
\label{sec:related_works}

\subsection{Species recognition from camera trap images with YOLO}

Fine-grained classification of wildlife can refer to the classification of individual species, whereas coarse-grained classification can refer to that of larger groups of species, such as functional groups, genus-level or family-level groups. In the past decade, there have been an increasing trend in using neural networks for image processing in ecology \cite{christin2019applications, nakagawa2023rapid}. Multiple studies have applied various iterations of YOLO to detect species of wildlife from Africa \cite{leonid2023human, schneider2018deep}, Europe \cite{schneider2018deep}, Asia \cite{tan2022animal}, and Australia \cite{nguyen2023sawit, falzon2019classifyme}, showing performance as good as those trained with two-stage models like Faster R-CNN in some instances.

However, Zurita et al. \cite{zurita2023use} noted that model performance was not ideal when trying to detect and classify two similar Amazonian pig species - white-lipped peccary (\textit{Tayassu pecari}) and collared peccary (\textit{Dicotyles tajacu}). Further, Schneider et al. \cite{schneider2018deep} demonstrated that the accuracy of species were especially low (0\% in some species) when there were insufficient images.

\subsection{Modelling based on functional groups}

Species-specific models are common in ecological studies because different species occupy different spatial, temporal, and ecological niches. These models are useful not only when studying the distribution of endangered species, but are also suitable where there may be bias towards patterns created by common species \cite{secco2024identifying}. 

Functional groups are groups of species that share similar ecosystem functions, such as diet, geographic niche, and life histories. Modelling based on functional or taxonomic groups in ecosystems is advantageous as some wildlife management strategies are based on larger groups, such as kangaroo harvesting \cite{2022-kangaroo-harvest-management-plan} and predator management \cite{prugh2019designing}.

Grouping species into coarser categories can also be particularly useful where data for a particular species is insufficient. Thus, incorporating information of another species with similar ecosystem functions could improve the predictability of their ecology and movement patterns. 

%% file: sec/3_methods.tex
\section{Methods}
\label{sec:methods}

\subsection{Wild Deserts dataset}

The Wild Deserts project is a reintroduction project based in Sturt National Park, near the north-western border of New South Wales, Australia. The project aims to reintroduce regionally extinct small mammals, such as the greater bilby (\textit{Macrotis lagotis}) and the Shark Bay bandicoot (\textit{Perameles bougainville}), back onto the landscape and monitor their activities where native and non-native predators are present. These predators include dingoes, cats, and foxes, all of which play important roles influencing the behaviour of other wildlife in the same ecosystem. Camera traps were placed on paths where wildlife were known to traverse in order to collect more images. The dataset consisted of 15000 images with 15 classes of wildlife collected from 30 different cameras around the study site. 

\subsection{Data labelling}

Ecologists from the Wild Deserts project selected images where wildlife were visible and provided class labels in a spreadsheet. We used MegaDetector 5.0 \cite{beery2019efficient} to filter and remove non-animal images, and to draw the bounding boxes around the individual animals with the confidence parameter set to 0.9. Afterwards, we imported the MegaDetector results into Label Studio to conduct a quality check on the bounding boxes, while matching class labels to each box.

The raw data was very imbalanced due to the natural distribution of species, with the most prominent class (Red kangaroo) having almost 700 times more instances than the least prominent class (Lizard), which is a common issue faced in ecological datasets. We, therefore, removed around 8000 Red kangaroo images, but otherwise left the original class distribution intact. We also removed images where the species label could not be ascertained. After cleaning, the final dataset had 6140 images in total. We split the images temporally with a 70:15:15 ratio to prevent the same sequences from being distributed across the training, validation, and testing sets. 

\begin{figure*}
  \centering
   \includegraphics[width=1.0\linewidth]{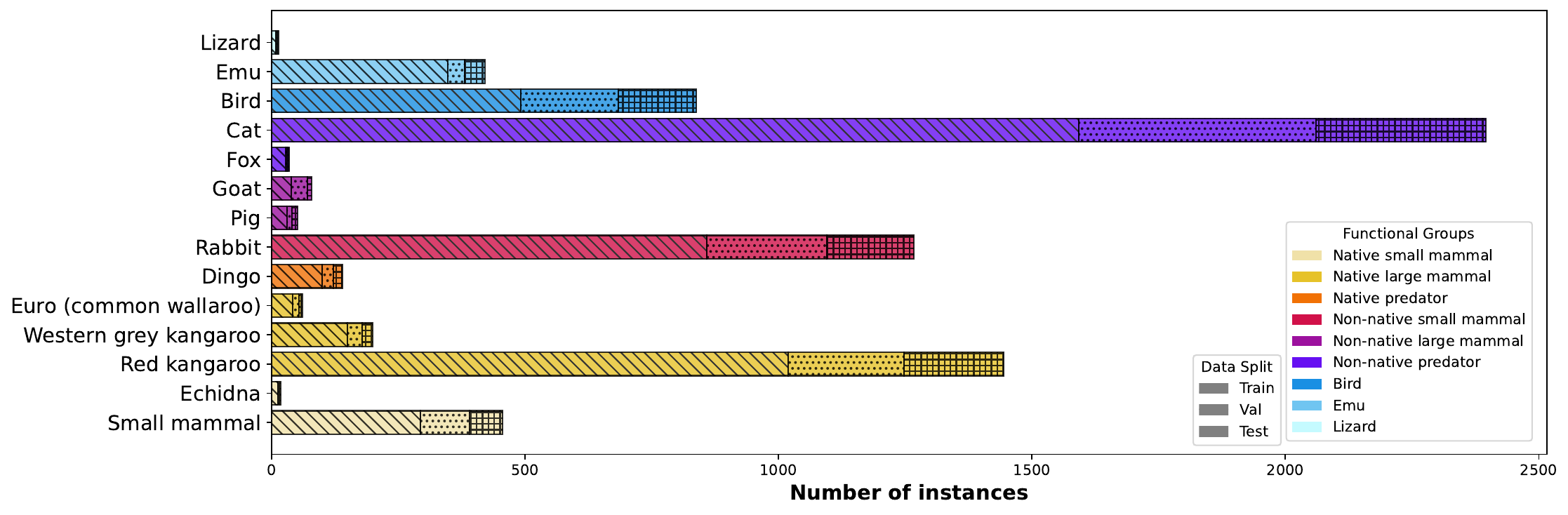}
   \caption{The distribution of the Wild Deserts dataset and the number of instances of each species.}
   \label{fig:one}
\end{figure*}

The fine-grained models consisted of 14 classes of wildlife, and the coarse-grained models had 9 classes (\cref{fig:one}). The groups in the coarse-grained models were formed by grouping certain species that had similar ecological functions. For instance, cats (\textit{Felis catus}) and foxes (\textit{Vulpes vulpes}) were grouped together as non-native predators since they are both predators that prey on small wildlife (e.g. small mammals, lizards, and birds), and were both introduced to Australia.

To test if empty background images could improve model performance, we added images that were absent of any wildlife by extracting around 20 images from each of the 30 cameras, which totalled to around 700 images (approximately 10\% of the dataset). We purposefully chose images that had different lighting and weather conditions.



\subsection{Metrics}

We report Mean Average Precision (mAP) at IoU=0.5 (AP50) for all experiments. For the fine-grained model, in additional to reporting AP50 for the set of fine-grained classes that the model was trained with, we also evaluate the ``coarse-grained'' AP50 by grouping fine-grained predictions according to our set of coarse-grained classes, and then evaluating this compared to coarse-grained ground truth using pycocotools \cite{pycocotools}.

\subsection{Training and testing}

The models in this project were ran with Python 3.11.4 \cite{python311}, CUDA 12.1 \cite{cuda}, and ultralytics 8.1.9 with YOLOv8 small\cite{yolov8_ultralytics}. All of the models were trained with YOLOv8 small with the addition of an early stopping policy where training was stopped 5 epochs after no improvements were made in training.

%% file: sec/4_results.tex
\section{Results}

\begin{figure*}
  \centering
   \includegraphics[width=0.49\textwidth]{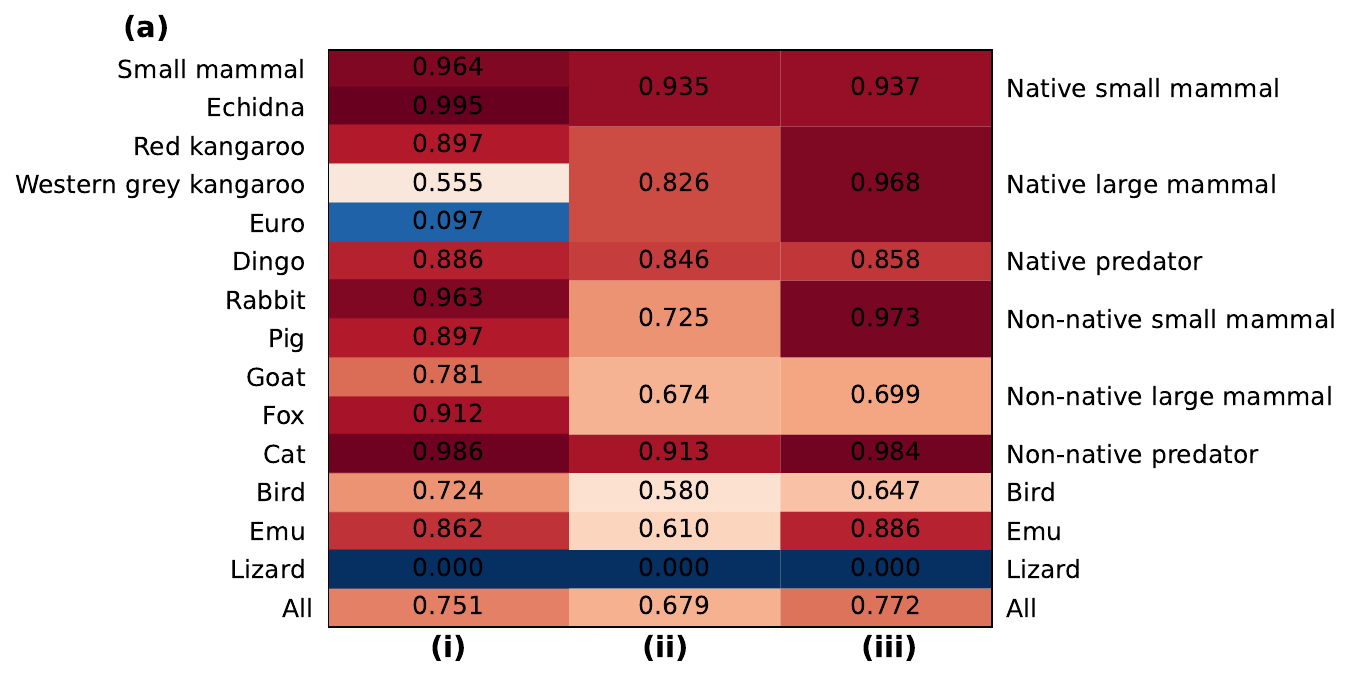}
   \includegraphics[width=0.49\textwidth]{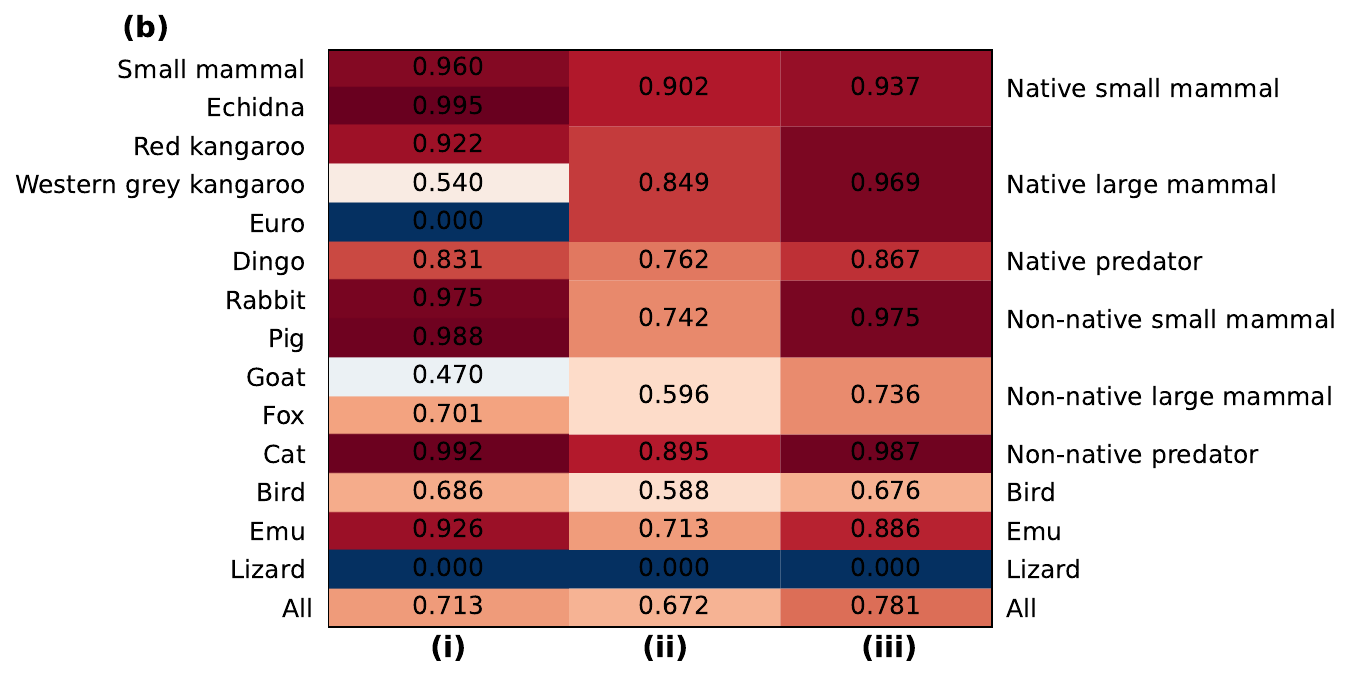}
   \includegraphics[width=0.4\textwidth]{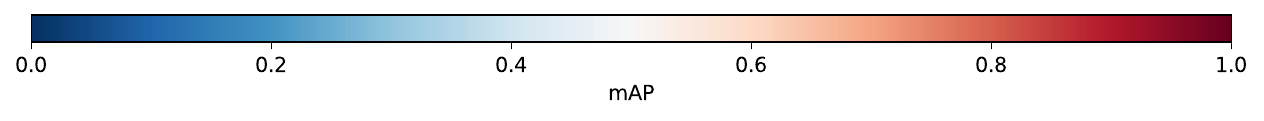}
   \caption{Mean average precision (mAP) of models (a) without and (b) with negative samples. Columns represent (i) mAP of the fine-grained classes from the fine-grained model calculated by YOLOv8, (ii) coarse-grained mAP extracted from the fine-grained model with pycocotools, and (iii) coarse-grained mAP from the coarse-grained model calculated by YOLOv8.}
   \label{fig:two}
\end{figure*}

We found that, on average, coarse-grained models performed slightly better than fine-grained models. Comparing the class-wise difference, coarse-grained models improved the mAP of some classes that had lower scores, while retaining most of the performance of other classes from the fine-grained models (\cref{fig:two}). 

\begin{figure*}
  \centering
   \includegraphics[width=0.32\textwidth]{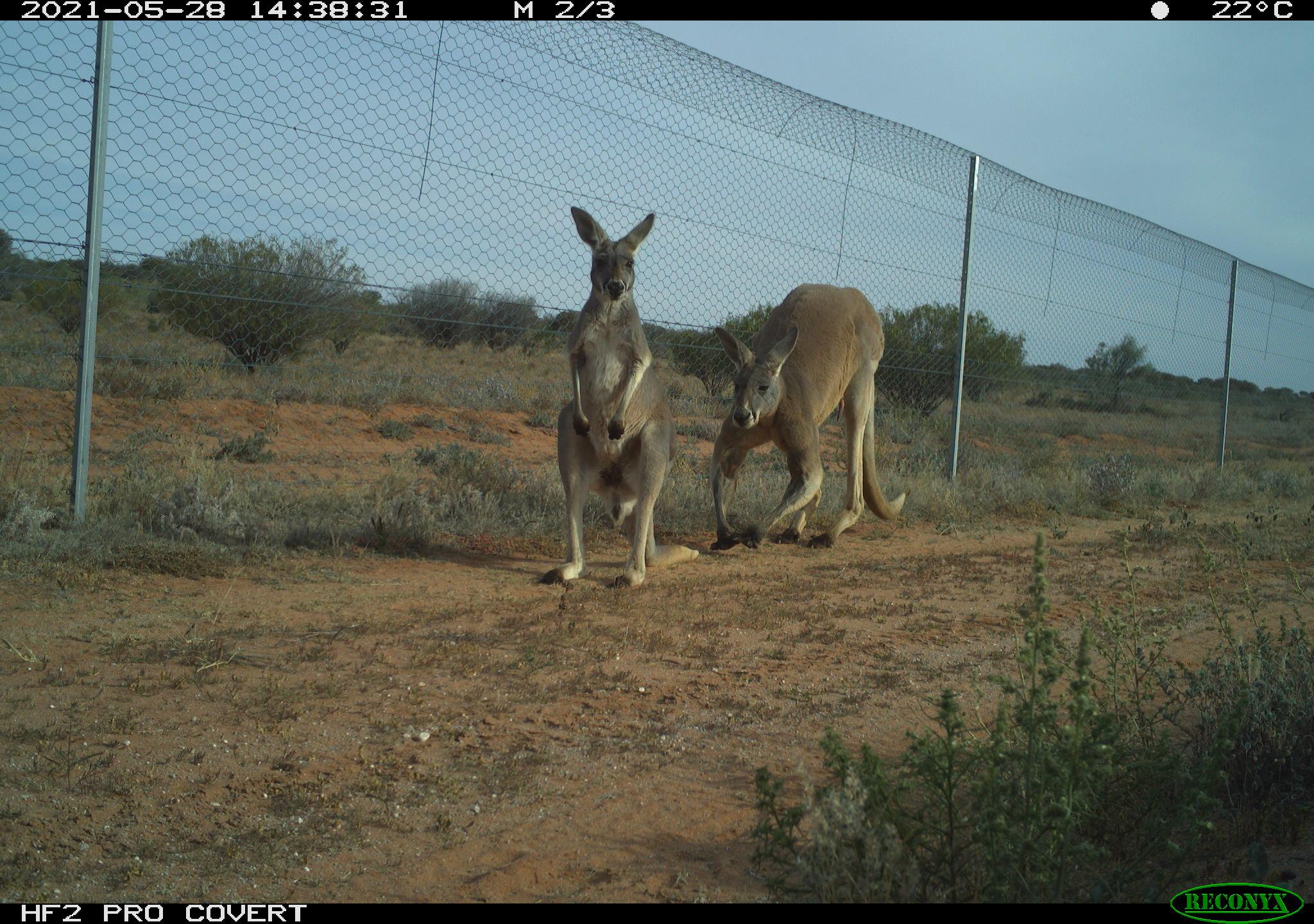} %
   \includegraphics[width=0.32\textwidth]{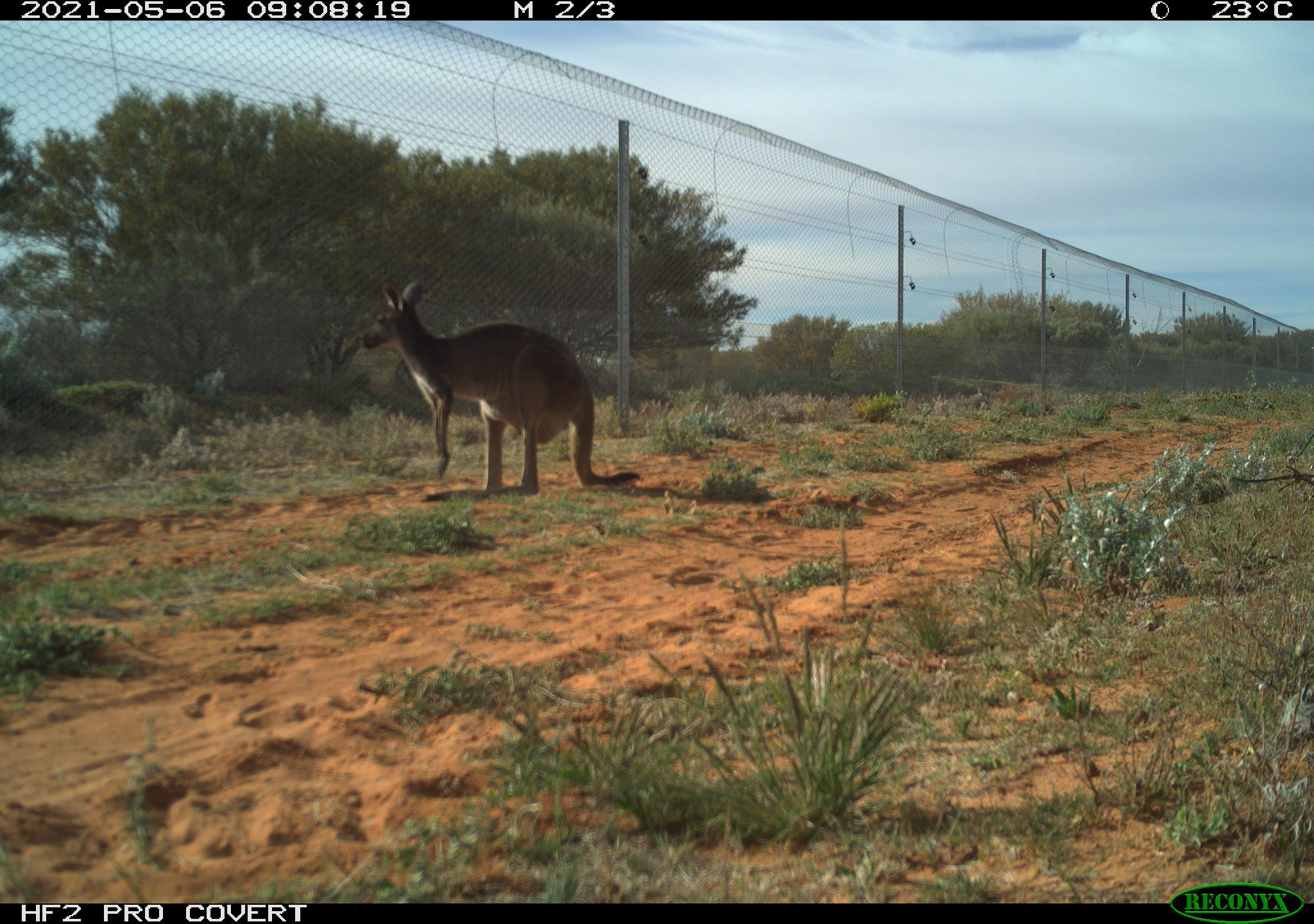} %
   \includegraphics[width=0.32\textwidth]{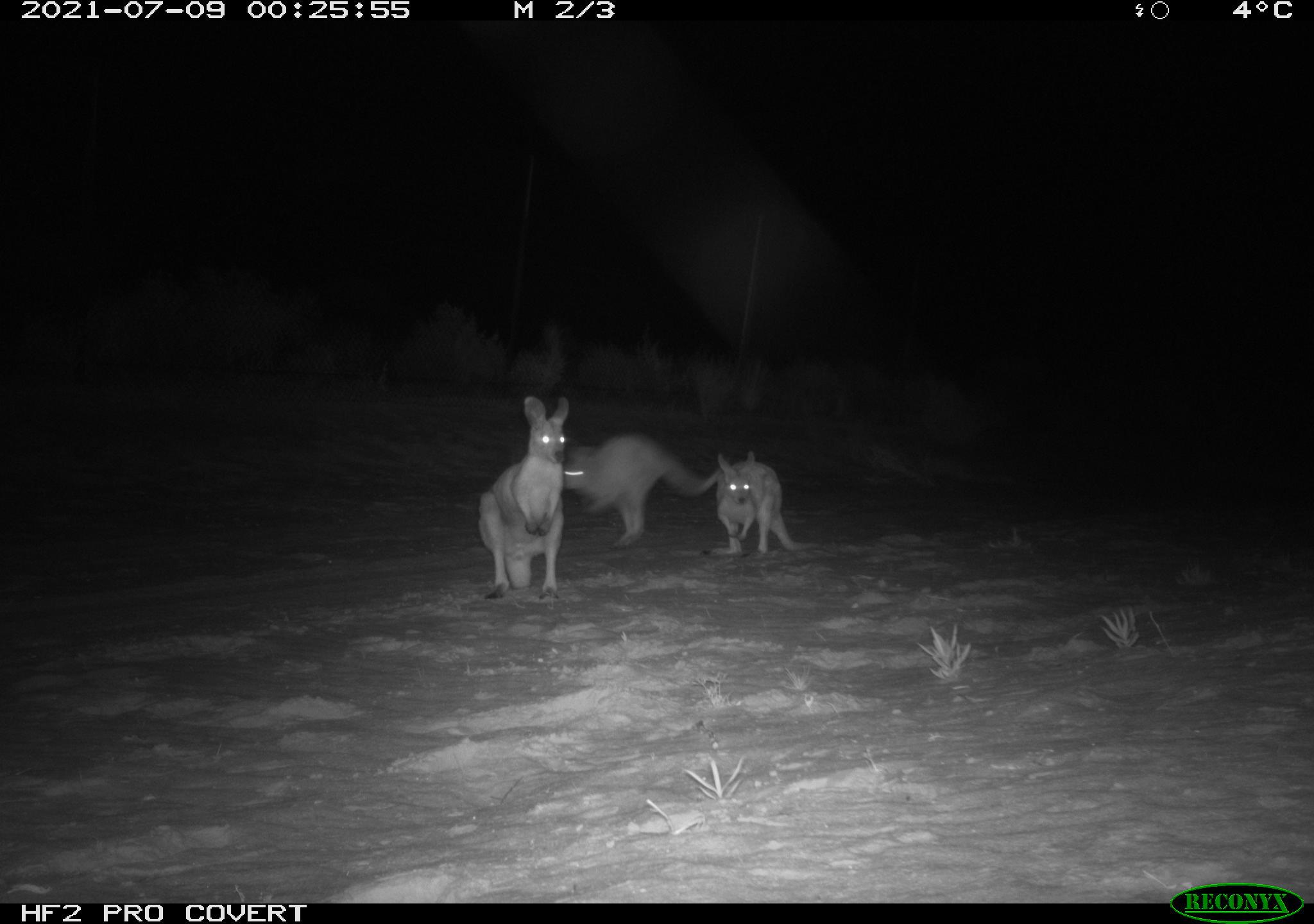}%
   \caption{Examples of macropod species from the Wild Deserts dataset. From the left: Red kangaroo (\textit{Osphranter rufus}), Western grey kangaroo (\textit{Macropus fuliginosus}), and Euro (\textit{Osphranter robustus})}
   \label{fig:three}
\end{figure*}

\begin{figure}
  \centering
   \includegraphics[width=1.0\linewidth]{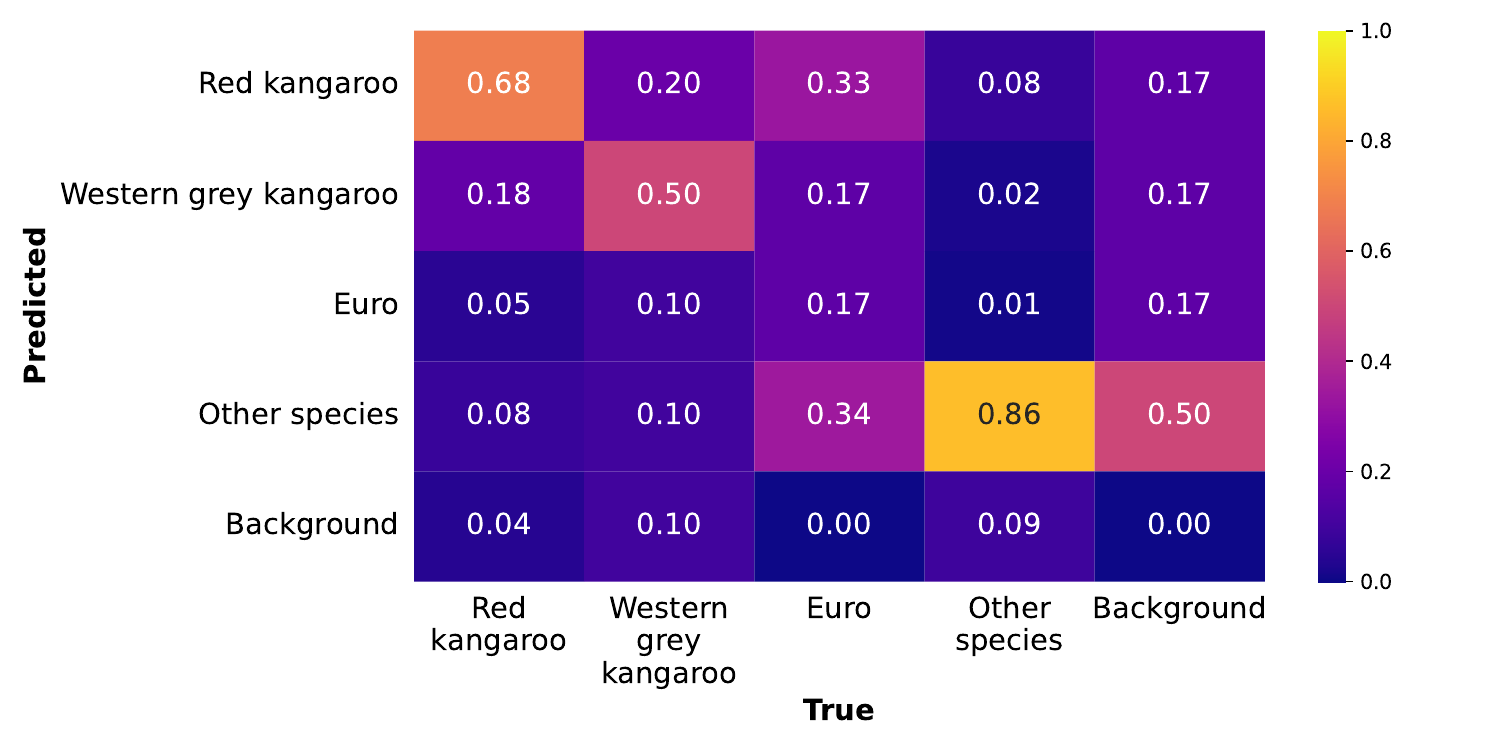}
   \caption{Normalised confusion matrix of the fine-grained model before adding negative samples.}
   \label{fig:four}
\end{figure}

Individual classes that gained the most performance from the coarse-grained models were Red kangaroo, Western grey kangaroo, and Euro (\cref{fig:three}). The mAP of all three macropod classes improved from 0.897, 0.555, and 0.097 respectively to 0.968 as Native large mammals without negative samples, and from 0.922, 0.540, and 0.000 to 0.969 with their addition. Euro especially benefited the most as only 17\% of the test images were correctly classified prior to the merge (\cref{fig:four}). Around one-third of Euro's test set was falsely identified as Red kangaroo and another one-third being other species.

The classes that had their mAP lowered when merged were Pig and Goat, from 0.897 and 0.781 to 0.699 as Non-native large mammals before the addition of negative samples. After the addition, their scores averaged out from 0.988 and 0.470 to 0.736.

On closer inspection, we observed some mixed results when comparing the mAP of the two fine-grained model, where adding negative samples to the training samples both improved and worsened the performance of models. For example, Pig detection improved by almost 0.1 mAP, while the detection of Goats dropped by 40\%. 

On the other hand, the improvements between the two coarse-grained models were marginal. The biggest difference we could observe was the improvement in non-native large mammals with the addition of negative samples, which increased by 0.037 mAP. 

%% file: sec/5_discussion.tex
\section{Discussion}
\label{sec:discussion}

Overall, we found that coarse-grained models performed better than fine-grained models especially for species with similar physical features, \textit{i.e.} Red kangaroo, Western grey kangaroo, and Euro. Whereas species that were not morphologically similar benefited less from being merged. Adding negative samples also saw an improvement in model performance to a certain extent.

\subsection{When to merge classes}

Although fine-grained models are useful in some contexts, distinguishing between morphologically similar species is difficult. The species-level mAP of macropods were lower than their combined functional group-level mAP for the same model, highlighting that recognising these species from one another was a challenging task for the model to accomplish. Some of the macropods' distinguishing features are fur colour, relative body size, and ear size. However, these characteristics can be difficult to observe in camera trap images especially when the images are less comprehensible due to factors such as lighting conditions and obstructed views \cite{Beery_2018_ECCV}. Leaving them as separate classes could increase false positive detections, especially if one class has significantly more training data than the others. Whereas species that look distinctively different and have sufficient data could be left as single classes.

\subsection{The inclusion of negative samples}

We included camera trap images without wildlife in the training set, but found mixed results. While we purposefully chose images from a variety of lighting situations, it may be that certain species only emerged during a specific time of the day, especially during the daytime, and that there were insufficient background images to represent those situations. Day images show more diverse ranges of colours and colour temperatures depending on the time of day and weather. For example, images tend to be warmer when the weather is sunny, but cooler when cloudy or rainy. Whereas night images are exclusively in black and white as they are captured by the camera's infrared technology.

\subsection{Other factors affecting model performance}

In some cases, class groupings were not enough to improve accuracy of rare classes.
For instance, the Lizard class, which was not part of any larger functional group, had a low score of 0 mAP in all models. This can be explained by the fact that there were very little observations to begin with (13 images in total). Following the same logic, the performance of Western grey kangaroo and Euro could potentially be improved as single classes by collecting more images, which could decrease false positives. 

\section{Conclusion}

While most existing studies focused on classifying individual species from camera traps, we provided an investigation into the differences between fine-grained and coarse-grained recognition of wildlife. We suggest that species that have similar physical features be grouped together, and species that look more distinctive need not be combined with others, given there is sufficient images. Whether to use fine-grained or coarse-grained models depend highly on the ecological question at hand. Further, we provided modest insight that the inclusion of negative samples could influence results, although a more comprehensive study will be required to investigate the threshold of empty images that could significantly benefit detection results.

%% file: sec/6_acknowledgement.tex
\section*{Acknowledgements}
\label{sec:Acknowledgements}

This work was made possible with the support of the Resnick Sustainability Institute and Computer Vision for Ecology summer workshop held at the California Institute of Technology. We also thank the ecologists from the UNSW Wild Deserts project for providing the labelled camera trap images.